\documentclass[preprint,preprintnumbers,amsmath,amssymb,superscriptaddress,nofootinbib]{revtex4}

\usepackage{graphicx}
\usepackage{psfrag}
\usepackage{axodraw}
\usepackage{subfigure}
\usepackage{color}
 \newcommand{\be}{\begin{equation}}
 \newcommand{\ee}{\end{equation}}
 \def\bea{\begin{eqnarray}}
 \def\eea{\end{eqnarray}}
 \newcommand{\f}[2]{\frac{#1}{#2}}

\begin{document}

\preprint{DCPT/06/136;  DESY 06-188;  IPPP/06/68; UAB-FT-612}

\author{Joerg Jaeckel}
\affiliation{Centre for Particle Theory, Durham University, Durham, DH1 3LE, UK}
\affiliation{Deutsches Elektronen Synchrotron, Notkestrasse 85, 22607 Hamburg, Germany}
\author{Eduard Mass{\'o}}
\affiliation{Grup de F{\'\i}sica Te{\`o}rica and Institut de F{\'\i}sica d'Altes Energies
Universitat Aut{\`o}noma de Barcelona, 08193 Bellaterra, Barcelona, Spain}
\author{Javier Redondo}
\affiliation{Grup de F{\'\i}sica Te{\`o}rica and Institut de F{\'\i}sica d'Altes Energies
Universitat Aut{\`o}noma de Barcelona, 08193 Bellaterra, Barcelona, Spain}
\author{Andreas Ringwald}
\affiliation{Deutsches Elektronen Synchrotron, Notkestrasse 85, 22607 Hamburg, Germany}
\author{Fuminobu Takahashi}
\affiliation{Deutsches Elektronen Synchrotron, Notkestrasse 85, 22607 Hamburg, Germany}
\title{The Need for Purely Laboratory-Based\\ Axion-Like Particle Searches}
\begin{abstract}
The PVLAS signal has led to the proposal of many experiments searching for light bosons coupled to
photons. The coupling strength probed by these near future searches is, however, far from the
allowed region, if astrophysical bounds apply. But the environmental conditions for the production
of axion-like particles in stars are very different from those present in laboratories. We consider
the case in which the coupling and the mass of an axion-like particle depend on environmental
conditions such as the temperature and matter density. This can relax astrophysical bounds by
several orders of magnitude, just enough to allow for the PVLAS signal. This creates exciting
possibilities for a detection in near future experiments.
\end{abstract} \maketitle

\section{Introduction}
Recently the PVLAS collaboration has reported the observation of  a rotation of the polarization
plane of a laser propagating through a transverse magnetic field \cite{Zavattini:2005tm}. This
signal could be explained by the existence of a new light neutral spin zero boson $\phi$, with a
coupling to two photons \cite{Maiani:1986md,Raffelt:1987im}
\be {\cal L}_I^{(-)}=\f{1}{4M}\phi^{(-)} F_{\mu\nu}\widetilde{F}^{\mu\nu} \hspace{.5cm} \textrm{or}
\hspace{.5cm} {\cal L}_I^{(+)}=\f{1}{4M}\phi^{(+)} F_{\mu\nu}F^{\mu\nu} \label{int-}\ee
depending on the parity of $\phi$, related to the sign of the rotation which up to now has not been
reported\footnote{The PVLAS collaboration has also found hints for an ellipticity
signal. The sign of the phase shift suggests an even particle $\phi^{(+)}$ \cite{PVLASICHEP}.}. Such an
Axion-Like Particle (ALP) would oscillate into
photons and vice versa in the presence of an electromagnetic field in a similar fashion as the
different neutrino
flavors oscillate between themselves while propagating in vacuum.

The PVLAS signal, combined with the previous bounds from the absence of a signal in the BFRT
collaboration experiment~\cite{Cameron:1993mr}, implies~\cite{Zavattini:2005tm}
\begin{equation}
1\ {\rm meV} \lesssim m \lesssim 1.5\ {\rm meV},
 \hspace{1.5cm}
2 \times 10^5\ {\rm GeV} \lesssim M
\lesssim 6\times 10^5\ {\rm GeV},
\label{masscoupPVLAS}
\end{equation}
with $m$ the mass of the new scalar.

It has been widely noticed that the interaction \eqref{int-} with the strength
\eqref{masscoupPVLAS} is in \textit{serious} conflict with astrophysical constraints
\cite{Raffelt:2005mt,Ringwald:2005gf}, while it is allowed by current laboratory and accelerator
data \cite{Masso:1995tw,Kleban:2005rj}. This has motivated recent work on building models that
evade the astrophysical constraints
\cite{Masso:2005ym,Jain:2005nh,Jaeckel:2006id,Masso:2006gc,Mohapatra:2006pv}, as well as
alternative explanations to the ALP hypothesis \cite{Antoniadis:2006wp,Gies:2006ca,Abel:2006qt}.

At the same time, many purely laboratory-based experiments have been proposed or are already on the way
to check the
particle interpretation of the PVLAS signal~\cite{Ringwald:2003ns,Rabadan:2005dm,Pugnat:2005nk,%
Gastaldi:2006fh,Afanasev:2006cv,Kotz:2006bw,Cantatore:Patras,BMV,Chen:2003tp,Gabrielli:2006im}. It
is important to notice, for the purpose of our paper, that these experiments are optical, and not
high-energy, accelerator experiments.

Quite generally, these experiments will have enough
sensitivity to check values of $M$ equal or greater than $10^6$
GeV, but, apart from Ref.~\cite{Ringwald:2003ns}, they do not have
the impressive reach of the astrophysical considerations, implying
$M\gtrsim 10^{10}$ GeV. Thus, if the PVLAS signal is due to
effects other than $\phi-\gamma$ oscillations and the
astrophysical bounds are applicable, these experiments can not
detect any interesting signal.

However, the astrophysical bounds rely on the assumption that the vertex \eqref{int-} applies under
typical laboratory conditions as well as in the stellar plasmas that concern the astrophysical
bounds. It is clear that, if one of the future dedicated laboratory experiments eventually sees a
positive signal, \textit{this can not be the case}.

In this work we investigate the simplest modification to the standard picture able to accommodate a
positive signal in any of the forthcoming laboratory experiments looking for ALPs, namely that the
structure of the interaction \eqref{int-} remains the same in both environments, while the values of
$M$ and $m$ can be different. Interestingly enough, the environmental conditions of stellar plasmas
and of typical laboratory experiments are very different and thus one could expect a very big
impact on $M$ and $m$.

We consider qualitatively the situation in which the dependence of $M$ and $m$ on the environmental
parameters produces a \textit{suppression} of ALP production in stellar plasmas. The main work of
the paper is devoted to compute this suppression using a realistic solar model and to investigate
how it relaxes the astrophysical bounds on the coupling \eqref{int-}. This leaves
 room for the proposed  laboratory experiments to potentially discover such an
axion-like particle.

In section \ref{sectionastrobounds} we revisit the astrophysical bounds and discuss
general mechanisms to evade them. In the
following section \ref{numerical}, we present our scenario of
environmental suppression and calculate the modified bounds. We
present our conclusions and comment on the reach of
proposed future laboratory experiments in section \ref{summary}.

\section{Astrophysical Bounds and General Mechanisms to Evade them\label{sectionastrobounds}}

\begin{figure}
  \includegraphics[width=6cm]{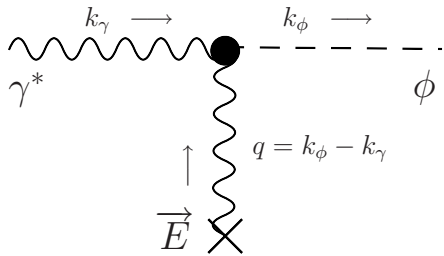}
  \caption{\label{fig0} Primakoff processes in which a photon turns into an
  ALP in the electric field of a charged particle like a proton or electron.}
\end{figure}

Presuming the $\phi\gamma\gamma$ vertex \eqref{int-}, photons of stellar plasmas can convert into
ALPs in the electromagnetic field of electrons, protons and heavy ions by the Primakoff effect,
depicted schematically in Fig.~\ref{fig0}. If $M$ is large enough, these particles escape from the
star without further interactions constituting a non-standard energy-loss channel. This energy-loss
channel accelerates the consumption of nuclear fuel and thus shortens the duration of the different
stages of stellar evolution with respect to the standard evolution in which ALPs do not exist.

In general, the astrophysical observations do agree with the theoretical predictions without
additional energy-loss channels so one is able to put bounds on the interaction scale $M$
\cite{Raffelt:1996wa}. The most important for our work are those coming from the lifetime of the
Sun \cite{Frieman:1987ui}, the duration of the red giant phase, and the population of Helium
Burning (HB) stars in globular clusters \cite{Raffelt:1985nk,Raffelt:1987yu}. The last of them
turns out to be the most stringent, implying
\be M > 1.7\times  10^{10} \ \textrm{GeV} \equiv M_{\rm HB}, \label{HBconstraint}\ee
for $m<{\mathcal O}(1\ {\rm keV})$. Moreover, if ALPs are emitted from the Sun one may try to
reconvert them to photons at Earth by the inverse Primakoff effect exploiting a strong magnetic
field. This is the helioscope idea \cite{Sikivie:1983ip} that it is already in its third generation
of experiments. Recently, the CERN Axion Solar Telescope (CAST) collaboration has published their
exclusion limits \cite{Zioutas:2004hi} from the absence of a positive signal,
\be M > 8.6\times 10^{9}\ \textrm{GeV}\equiv M_{\rm CAST}, \label{CASTconstraint}\ee
for $m<0.02$~eV.

One should be aware that these astrophysical bounds rely on many assumptions to calculate the flux
of ALPs produced in the plasma. In particular, it has been assumed widely in the literature that
the same value of the coupling constant that describes $\phi-\gamma$ oscillations in a magnetic
field in vacuum describes the Primakoff production in stellar plasmas, and the mass has been also
assumed to be the same. We want to remark that this has been mainly an argument of pure simplicity.
In fact, there are models in which $M$ depends on the momentum transfer $q$ at which the vertex is
probed \cite{Masso:2005ym} or on the  effective mass $\omega_P$ of the plasma photons involved
\cite{Masso:2006gc}. These models have been built with the motivation of evading the astrophysical
bounds on  ALPs, by decreasing the  effective  value of the coupling $1/M$ in stellar plasmas in
order to solve the inconsistency between the ALP interpretation of PVLAS and the astrophysical
bounds. This has proven to be a very difficult task because of the extreme difference between the
PVLAS value (\ref{masscoupPVLAS}) and the HB (\ref{HBconstraint}) or CAST (\ref{CASTconstraint})
exclusion limits. These models require very specific and somehow unattractive features like the
presence of new confining forces or tuned cancellations (note, however, \cite{Abel:2006qt}).
Anyway, they serve as examples of how $M$ (and eventually $m$) can depend  on ``environmental''
parameters $\eta = q,\omega_P$, etc... (for other suitable parameters, see
Table~\ref{PVLAS_SUN_HB}),
\begin{equation}
M\rightarrow M(\eta), \quad m\rightarrow m(\eta),
\end{equation}
such that the production of ALPs is suppressed in the stellar environment.

\begin{table}
\begin{tabular}{c|c|c|c}
Env. param. &  Solar Core & HB Core & PVLAS \\ \hline
$T$ [keV] &  $1.3$ & $8.6$ & $\sim 0$
\\ \hline
$q^2$ [keV$^{2}$] & $\sim 1$ & $\sim 1$ & $\sim 10^{-12}$ \\
\hline $\omega_P$ [keV] & $0.3$ & $2$ & $0$ \\
\hline $\rho$ [g~cm$^{-3}$] & $1.5\times 10^2$& $10^4$ & $< 10^{-5}$ \\
\end{tabular}
\caption{\label{PVLAS_SUN_HB}
Comparison between the values of environmental parameters, such as
the temperature $T$, typical momentum transfer $q$, plasma frequency $\omega_P$, and
matter energy density $\rho$, in the stellar plasma and in the PVLAS experiment.
Other parameters to consider could be the Debye screening scale $k_s$,
or, to name something more exotic, the neutrino flux, or  the average
electromagnetic field.}
\end{table}

\begin{figure}
  \begin{center}
\includegraphics[width=10.5cm]{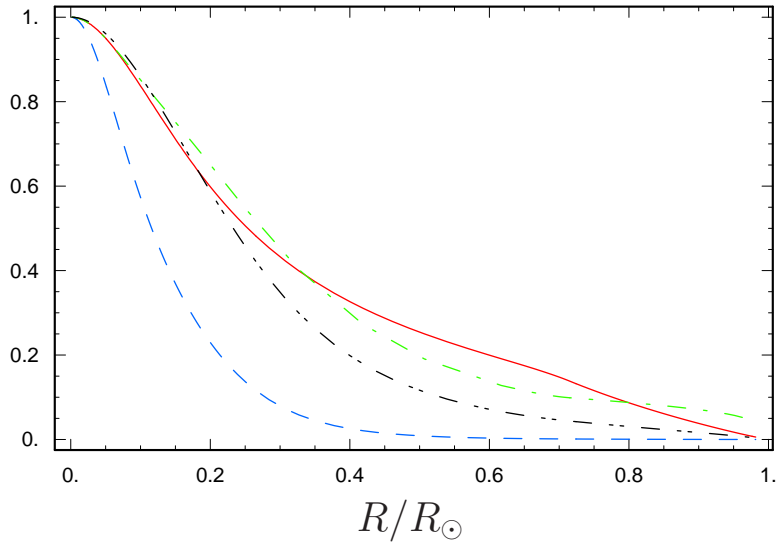}
\end{center}
\vspace{-1cm}
  \caption{\label{fig2} Environmental parameters as a function of the distance to the solar center.
Temperature (solid, red), matter density (dashed, blue),
  Debye screening scale (double dashed, green) and plasma frequency
  (triple dashed, black), normalized to their values in the solar center,
  $T_0=1.35$ keV, $\rho_0=1.5\times 10^{2}$~g\,cm$^{-3}$, $k_{s0}=9$~keV, $\omega_{P0}=0.3$~keV
  for the solar model BS05(OP) of Bahcall {\it{et al.}} \cite{Bahcall:2004pz}.
  }
\end{figure}

In the following, we will not try to construct micro-physical explanations for this dependence but
rather write down simple effective models and fix their parameters in order to be consistent with
the solar bounds and PVLAS or any of the proposed laboratory experiments.

A suppression of the production in a stellar plasma could be realized in
two simple ways:
\begin{itemize}
\item[(i)] either the coupling $1/M$ decreases (dynamical suppression) or
\item[(ii)] $m$ increases
to a value higher than the temperature such that the production is Boltzmann suppressed
(kinematical suppression).
\end{itemize}

All the environmental parameters considered in this paper are  \textit{much} higher in the Sun than
in laboratory conditions (see Table~\ref{PVLAS_SUN_HB} and Fig.~\ref{fig2}), so we shall consider
$M(\eta)$ and $m(\eta)$ as monotonic increasing functions of $\eta$ with the values of $M(\sim 0)$
and $m(\sim 0)$ fixed by the laboratory experiments.

Clearly, both mechanisms are efficient at suppressing the production of ALPs in the Sun, but there
is a crucial difference that results in some prejudice against  mechanism (ii). Mechanism (i) works
by making the already weak interaction between ALPs and the photons even weaker. The second
mechanism, however, is in fact a strong interaction between the ALPs and ordinary matter, thereby
making it difficult to implement without producing unwanted side effects. We will nevertheless
include mechanism (ii) in our study, but one should always keep this caveat in mind.

As we said, $\eta$ in the stellar plasma is generally much higher than in
laboratory-based experiments. It is then possible that
new ALP physics produces  also a big
difference between the values of the ALP parameters, $m$ and $M$, in such different
environments.

\begin{figure}[t]
  \centering
   \includegraphics[height=3.3cm]{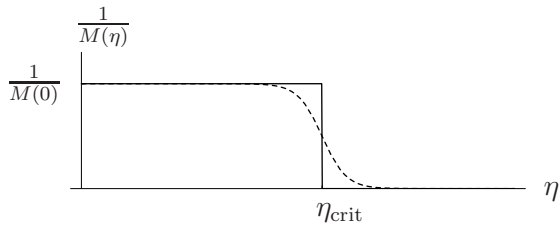}\vspace{-.5cm}
  \caption{\label{fig1} Coupling as a function of an environmental parameter $\eta$:
The simple form used in our calculations (solid line) and a generic,
more realistic, dependence (dashed).}
\end{figure}

Let us remark on the  a priori unknown shape of $M(\eta)$ and $m(\eta)$. In our calculations we use
a simple step function (cf. Fig.~\ref{fig1}), which has only one free parameter: the value for the
environmental parameter where the production is switched off, $\eta_{\rm{crit}}$. In most
situations this will give the strongest possible suppression. The scale $\eta_{\rm{crit}}$ can be
associated with the scale of new physics responsible for the suppression. In what follows, we will
consider only the effects of one environmental parameter at once although it is trivial to
implement this framework for a set of parameters.

For simplicity, we restrict the study of the environmental suppression of ALPs to our Sun because
we know it quantitatively much better than any other stellar environment. The group of Bahcall has
specialized in the computation of detailed solar models which provide all the necessary ingredients
to compute accurately the Primakoff emission. We have used the newest model, BS05(OP)
\cite{Bahcall:2004pz}, for all the calculations of this work (our accuracy goal is roughly $10\%$).
The variation of some environmental parameters is displayed in Fig.~\ref{fig2} as a function of the
distance from the solar center.

\section{\label{numerical}Numerical Results}

Let us first state how a suppression $S$ of the flux of ALPs affects the bounds arising from energy
loss considerations and helioscope experiments. If the flux of ALPs from a stellar plasma is
suppressed by a factor $S$, the energy loss bounds on $M$  are relaxed by a factor of
$\sqrt{S}$ while the CAST bound relaxes with $\sqrt[4]{S}$,
\begin{eqnarray}
\label{modloss} M_{\rm{loss}} \rightarrow \sqrt{S}\,M_{\rm{loss}},&&\hspace{6ex} {\rm
energy\ loss\ bound},
\\
\label{modcast} M_{\rm{CAST}} \rightarrow \sqrt[4]{S}\,M_{\rm{CAST}},&&\hspace{6ex} {\rm{CAST\
bound}},
\end{eqnarray}
since the former depends only on the Primakoff production, $\sim 1/M^2$, and the latter gets an
additional factor $\sim 1/M^2$ for the reconversion at Earth  resulting in a total counting rate
$\sim 1/M^4$.

\subsection{Dynamical Suppression}

We consider first a possible variation of the coupling that we have enumerated as mechanism (i).
Treating the emission of ALPs as a small perturbation of the standard solar model, we can compute
the emission of these particles from the unperturbed solar data. The $\gamma-\phi$ Primakoff
transition amplitude can be written as (neglecting the plasma mass $\omega_{P}$ for the
moment)\footnote{We are using natural units $\hbar = c =1$ with the Boltzman constant, $k_B=1$.}
\be \label{Gammanoplasma}
\Gamma(\omega)_{\gamma-\phi}=
\frac{T k^{2}_{s}}{64\pi}
\int^{1}_{-1}\hspace{-7pt}d\cos\theta
\f{1+\cos\theta}{\kappa^2+1-\cos\theta}\f{1}{M(\eta)^2},\ee
where $\omega$ is the energy of the incoming photon, and
\begin{equation}
k^{2}_{s}=\frac{4\pi\alpha}{T} (n_{e}+\sum_{i}Z^{2}_{i}n_{i}),
\end{equation}
is the Debye screening scale. $n_i,Z_i$ are the number densities and charges of the different
charged species of the plasma, $\alpha\simeq 1/137$, $n_{e}$ is the electron number density,
$\cos\theta$ is the relative angle between the incoming photon and the outgoing ALP in the target
frame (considered with infinite mass) and $\kappa^2= k_s^2/2\omega^2$. Integration over the whole
Sun with the appropriate Bose-Einstein factors for the number density of photons gives the spectrum
of ALPs (number of emitted ALPs per unit time per energy interval),
\be \label{spectrum}\f{d^2N(\omega)}{d\omega dt}=4\pi\int_0^{R_\odot}\hspace{-9pt}R^2dR
\f{\omega^2}{\pi^2} \f{\Gamma(\omega)_{\gamma-\phi}}{{\rm e}^{\omega/T}-1}  \ \ \ .\ee
(Remember that $T$, $k_s^2$, etc. depend implicitly on the distance $R$
from the solar center.)

\begin{figure}
\begin{center}
  \includegraphics[width=10.5cm]{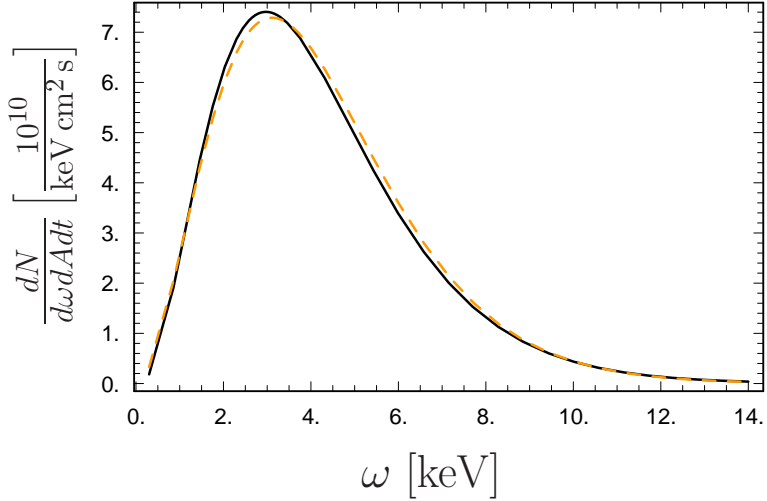}
\end{center}
\vspace{-1cm}
  \caption{\label{figextra} Our spectrum of ALPs at Earth (black solid) agrees reasonably well with that of
  the CAST collaboration \cite{Zioutas:2004hi}
  (dashed orange) for $M=10^{10}\,\rm{GeV}$.}
\end{figure}

As a check of our numerical computation we have computed the flux of \textit{standard} ALPs at
Earth which is shown in Fig.~\ref{figextra} and does agree with the CAST calculations
\cite{Zioutas:2004hi}.

It is very important to differentiate two possibilities:
\begin{itemize}
\item[A)] $\eta$ is a macroscopic (averaged) environmental parameter given by the solar
model and depending only on the distance $R$ from the solar center. Then the suppression acts as a
step function in the $R$ integration \eqref{spectrum} for the flux.
\item[B)] $\eta$ depends on the microscopic aspects of the production like the momentum transfer $q^2$.
Then the step function acts inside the integral in eq. \eqref{Gammanoplasma}.
\end{itemize}

We now start with the first possibility and let the second, which requires a different treatment,
for subsubsection \ref{sectionq2}.

\subsubsection{Dynamical suppression from macroscopic environmental parameters}\label{macroscopic}

If $1/M(\eta)$ is a step function, ALP production is switched off wherever $\eta>\eta_{\rm crit}$.
Let us call $R_{\rm crit}$ the radius at which the coupling turns off, i.e. $\eta(R_{\rm
crit})=\eta_{\rm crit}$. Since the functions $\eta(r)$ shown in Fig.~\ref{fig2} are monotonous, we
can calculate the suppression as a function of $R_{\rm crit}$ and then determine $\eta_{\rm
crit}=\eta(R_{\rm crit})$.

We define the suppression efficiency, $S(\omega,R_{\rm crit})$,  as the ratio of the flux of ALPs
with energy $\omega$ with suppression, divided by the one without suppression,
\begin{equation}
S(\omega;R_{\rm crit})=\frac{d^2N(\omega;R_{\rm crit})}{d\omega\,dt}
\left(\frac{d^2N(\omega)}{d\omega\,dt}\right)^{-1}.
\end{equation}
%

\begin{figure}
    \includegraphics[width=10.5cm]{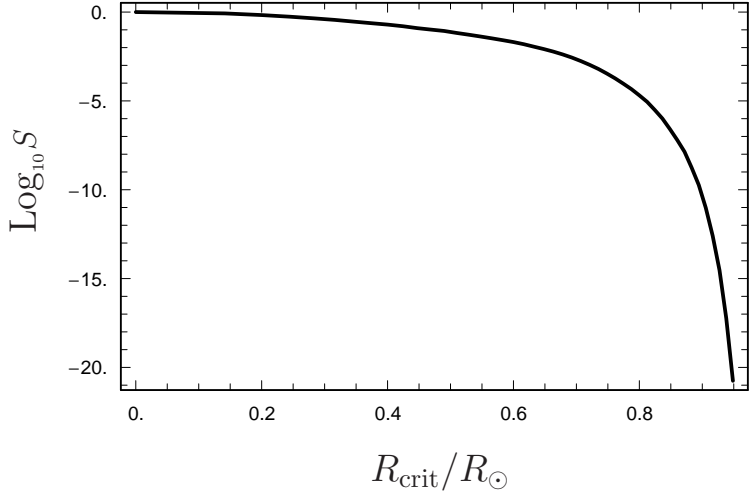} 
\vspace{-.5cm}
\caption{\label{fig4}Suppression of the flux of ALPs
$S(\omega_0=1\,\rm{keV},R_{\rm{crit}})$ as a function of $R_{\rm crit}$.}
\end{figure}

The CAST experiment is only sensitive to ALPs in the range of $(1-14)\,\rm{keV}$.
Hence, we must suppress the production of ALPs only in this energy range.
In order to provide a simple yet conservative bound
we use the factor $S(\omega_{0},R_{\rm{crit}})$ evaluated at the energy
$1\,\rm{keV}\leq \omega_{0}\leq 14\,\rm{keV}$ which maximizes $S$.
We have checked that, in all
cases of practical interest, $\omega_0$ is the CAST lower threshold, 1~keV.
In Fig.~\ref{fig4}, we plot $S(1\,{\rm keV},R_{\rm crit})$. In Tab. \ref{tab2} we give some values
for $S$ together with the corresponding values of $\eta_{\rm{crit}}$.

From the modified CAST bound (\ref{modcast}),
\be
 M > \sqrt[4]{S(\omega_0,R_{\rm crit})}\ 8.7\times 10^{9}\ \textrm{GeV},
 \label{CASTmodifiedconstraint}
\ee
we infer that in order to reconcile it with the PVLAS result, $M_{\rm PVLAS}\sim (10^5-10^6)$~GeV,
we need
\be S_{\rm CAST}\sim 10^{-20} \ \ \ . \label{castfactor} \ee
Looking at Table~\ref{tab2}, we find that this is possible, but the critical environmental
parameters are quite small; for example, the critical plasma frequency is in the eV range.
Moreover, the results  are sensitive to the region close to the surface of the Sun where $\log(S)$
changes very fast and our calculation becomes somewhat less reliable.

\begin{table}
\begin{tabular}{c|c|c|c|c}
$R_{\rm crit}/R_\odot$ & $T_{\rm crit}\,[{\rm keV}]$ &
$\rho_{\rm crit}\,[{\rm g}\, {\rm cm}^{-3}]$ & $\omega_{P,\rm{crit}}\,[{\rm keV}]$ & $S$\\
\hline $ 0$ & $1.35$ & $150$ & $0.3$ & $1$ \\
\hline $0.2$ & $0.81$ & $35$ & $0.16$ & $0.67$ \\
\hline $0.5$ & $0.34$ & $1.3$ & $0.03$ & $0.08$ \\
\hline $0.7$ & $0.2$ & $0.2$ & $0.01$ & $2\times 10^{-3}$ \\
\hline $0.8$ & $0.12$ & $0.09$ & $0.008$ & $ 2\times10^{-5}$ \\
\hline $0.85$ & $0.08$ & $0.05$ & $0.006$ & $2\times10^{-7}$ \\
\hline $0.9$ & $0.05$ & $0.03$ & $0.004$ & $4\times10^{-11}$ \\
\hline $0.95$ & $0.025$ & 0.009 & 0.0025 & $\sim 10^{-20}$\\
\end{tabular}
\caption{\label{tab2}Several values of $S(\omega_0=1\,\rm{keV},R_{\rm{crit}})$ with their
respective values of the suppression scales $\eta_{\rm crit}$.}
\end{table}

We now take a look at the solar energy loss bound (\ref{modloss}). The age of the Sun is known to
be around $5.6$ billion years from radiological studies of radioactive crystals in the solar system
(see the dedicated Appendix in \cite{Bahcall:1995bt}). Solar models are indeed built to reproduce
this quantity (among others, like today's solar luminosity, solar radius, etc...), so one might
think that a model with ALP emission can be constructed as well to reproduce this lifetime.
However, this seems not to be the case for large ALP luminosity \cite{Raffelt:1987yu} and it is
concluded that the exotic contribution cannot exceed the standard solar luminosity in photons. For
our purposes this means
\be L_{\rm ALP}<L_\odot=3.846\times10^{26}\ {\rm W} \sim 1.60\times 10^{30}\ {\rm eV}^2
\label{LifetimeBound},\ee
with
\be L_{\rm ALP} \equiv \int_0^\infty d\omega\, \omega\,\frac{d^2N}{d\omega dt}. \ee

We have computed the ALP emission in BS05(OP),
\be L_{\rm ALP}= 1.8\times 10^{-3} \left(\f{10^{10}\ \textrm{GeV}}{M}\right)^2 L_\odot .
\label{SunConstraint}\ee
This value is slightly bigger than that of Ref.~\cite{vanBibber:1988ge}, which relies on an older
solar model \cite{Bahcall:1981zh}, probably as a consequence of the different data.

\begin{figure}
    \includegraphics[width=10.5cm]{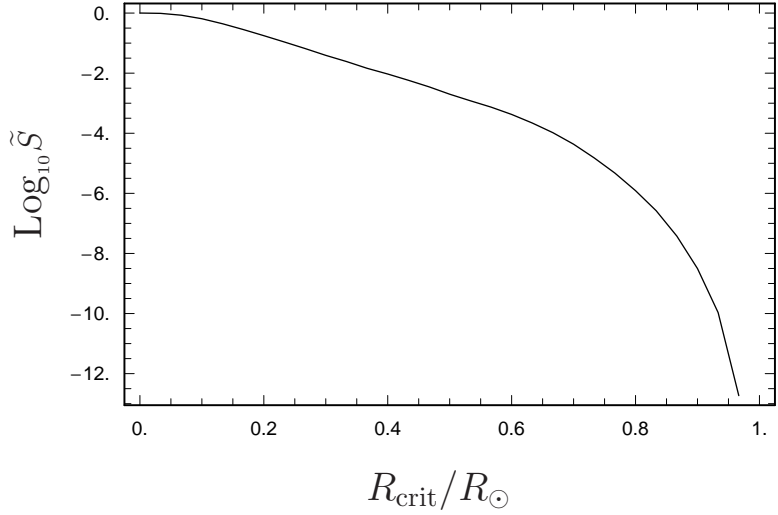} 
\caption{\label{totalflux1}Suppression $\tilde{S}$ of the total flux of ALPs as a function of the
critical radius $R_{\rm{crit}}$.}
\end{figure}

For the total flux, we find a suppression
\begin{equation}
\tilde{S}(R_{\rm{crit}})=\frac{L_{\rm{ALP}}(R_{\rm{crit}})}{L_{\rm{ALP}}}
,
\end{equation}
which we plot in Fig. \ref{totalflux1}. Using the modified energy loss bound \eqref{modloss},
\eqref{LifetimeBound} and \eqref{SunConstraint} we get
\begin{equation}
M>\sqrt{\tilde{S}(R_{\rm{crit}})}\ 6\times 10^9\ \textrm{GeV},
\end{equation}
and we need a much more moderate
\begin{equation}
\tilde{S}_{\rm{loss}}\sim 10^{-10}
\label{lossfactor}
\end{equation}
to avoid a conflict between the PVLAS result and the energy loss argument.
Accordingly, this bound alone requires values for the critical environmental parameters that are larger
(and therefore less restrictive) than those from the CAST bound.

\subsubsection{Dynamical suppression from microscopic parameters: $q^2$ \label{sectionq2}}

In the previous subsection, we have considered macroscopic environmental parameters
like, e.g., the temperature $T$. However, suppression could also result from a dependence on
microscopic parameters like, e.g., the momentum transfer $q^2$ in a scattering event (not averaged).

In this section we discuss the well motivated (cf. \cite{Masso:2005ym}) example of a possible
dependence $M=M(q^2)$ on the momentum transfer involved in
the Primakoff production (Fig.~\ref{fig0}).
Again, we use a step function to model the dependence on $q^2$,
\be \frac{1}{M(q^2)}=\frac{1}{M(0)}\ \Theta(q^{2}_{\rm{crit}}-|q^2|)
=\frac{1}{M(0)}\ \Theta(q^2_{\rm{crit}}-q_m^2-2k_\phi
k_\gamma(1-\cos\theta)) \label{Mq2stepfunction},
\ee
where $k_\phi$, $k_\gamma$ are the moduli of the momenta of the ALP and the photon.
$q_m=|k_\phi-k_\gamma|$ is the smallest possible momentum transfer. Here, we will use the
approximation $m=0$, but it will be crucial to take into account that
photons have an effective mass
\begin{equation}
m^{2}_\gamma=\omega^{2}_P=\frac{4\pi\alpha n_{e}}{m_{e}},
\end{equation}
so $q_m(\omega)=\omega-\sqrt{\omega^2-\omega_P^2}$. Note that the plasma mass is crucial because it
ensures that $q_m>0$, i.e. it removes ALP production processes with very small momentum transfer
which would be unsuppressed.

With this modification, Eq.~\eqref{Gammanoplasma} reads
\be \Gamma_{\gamma-\phi}(\omega)
=
\frac{T k^{2}_{s}}{64\pi}
\int_{-1}^{+1}d\cos\theta
\f{\sin^2\theta}{(x-\cos\theta)(y-\cos\theta)}\f{1}{M^2(q^2)} \label{gammaunsuppressed} \ee
with  $x=(k_\phi^2+k_\gamma^2)/2k_\phi k_\gamma$ and $y=x+k_s^2/2k_\phi k_\gamma$.
The step function implies that only values of $\cos\theta$ satisfying
\be \cos\theta > 1 - \f{q^2_{\rm{crit}}-q^2_m}{2\omega\sqrt{\omega^2-\omega^2_P}} \ee
contribute to the integral. Hence, we find that the effect of the step
function \eqref{Mq2stepfunction} is to
restrict the integration limits of Eq.~(\ref{gammaunsuppressed}),
\be \Gamma_{\gamma-\phi}(\omega)=\f{T k_s^2}{64\pi M^2(0)}\int_{\delta(\omega)}^{+1}d\cos\theta
\f{\sin^2\theta}{(x-\cos\theta)(y-\cos\theta)}, \label{gammasuppressed} \ee
with
\be \delta(\omega)= \left(
\begin{array}{cccc} 1 & \textrm{for} & q_{\rm{crit}} < q_m(\omega)& \\
 1 - \f{q_{\rm{crit}}^2-q^2_m(\omega)}{2\omega\sqrt{\omega^2-\omega^2_P}} & \textrm{for}
& q_{\rm{crit}} > q_m(\omega),&
1 - \f{q_{\rm{crit}}^2-q^2_m(\omega)}{2\omega\sqrt{\omega^2-\omega^2_P}}>-1
\\
                 -1 \hspace{1cm} & \textrm{for} & q_{\rm{crit}} > q_m(\omega),&
1 - \f{q_{\rm{crit}}^2-q^2_m(\omega)}{2\omega\sqrt{\omega^2-\omega^2_P}}\leqslant-1
\end{array} \right). \ee
When $\delta(\omega)=1$, the integral is zero and Primakoff conversion is completely suppressed.
This happens for values of the plasma frequency $\omega_P$ and the energy $\omega$ for which the
minimum momentum transfer is already larger than the cut-off scale $q_{\rm{crit}}$. We point out
that this is an energy dependent statement. For $\omega\gg\omega_P$ large enough, $q_m$ is small
enough to satisfy $q_{\rm{crit}}\gg q_m$. When this is the case we have only partial suppression.
The integral goes only over the small interval $[\delta(\omega),1]$ where $\delta(\omega)\approx
1-q_{\rm{crit}}^2/2\omega^2$,  $x\approx 1$ and
$y\approx 1+k_s^2/2\omega^2$. Then the integral can be easily estimated by the value of the
integrand at $\cos\theta=1$,
\be \Gamma_{\gamma-\phi}(\omega)\sim \f{T k_s^2}{64\pi M^2}
\f{4\omega^2}{k_s^2}\f{q_{\rm{crit}}^2}{2\omega^2},
\quad\quad\quad {\rm{for}}\,\, \omega\gg \omega_P,\, k_{s},\,q_{\rm{crit}}.
\ee
Notice that although we have used the strongest possible suppression,
a step function, at the end of the day, at high energies, the transition rate is only suppressed by a factor
$q_{\rm{crit}}^2/k_s^2$. This means that the $\gamma^*
-\phi$ transition is suppressed at most quadratically.

This holds even for a generic
suppressing factor 
$F(q^2) = M(q^2)/M(0)$.
The limitation comes from the part of the integral which
is close to $\cos\theta=1$. There the integrand is a constant, $1+\cos\theta/(y-1)\sim
4\omega^2/k_s^2$. By continuity, the suppression factor $F(q^2)$, whatever it is, must be close to unity
because $q^2$ is very close to zero and normalization requires $F(q^2=0)=1$. This holds for values of
$q^2$ up to a certain range, limited by the shape of $F(q^2)$. Defining $q_{\rm{crit}}^2$ as the size of
the interval where $F(|q^2|\lesssim q_{\rm{crit}}^2)\sim 1$, then $q_{\rm{crit}}^2=|q^2|$ gives a minimum
value for $\cos\theta$ for which the integrand is nearly constant
($\cos\theta_m \sim 1-q_{\rm{crit}}^2/2\omega^2$), leading to
\be \Gamma(\omega)\propto \int^1_{-1}d\cos\theta\,
\f{1+\cos\theta}{y-\cos\theta}\,F(q^2)\gtrsim\int^1_{\cos\theta_m}d\cos\theta\,
\f{4\omega^2}{k_s^2}\sim \f{2q_{\rm{crit}}^2}{k_s^2}
.\ee
%

\begin{figure}
    \centering
    \includegraphics[width=7.5cm]{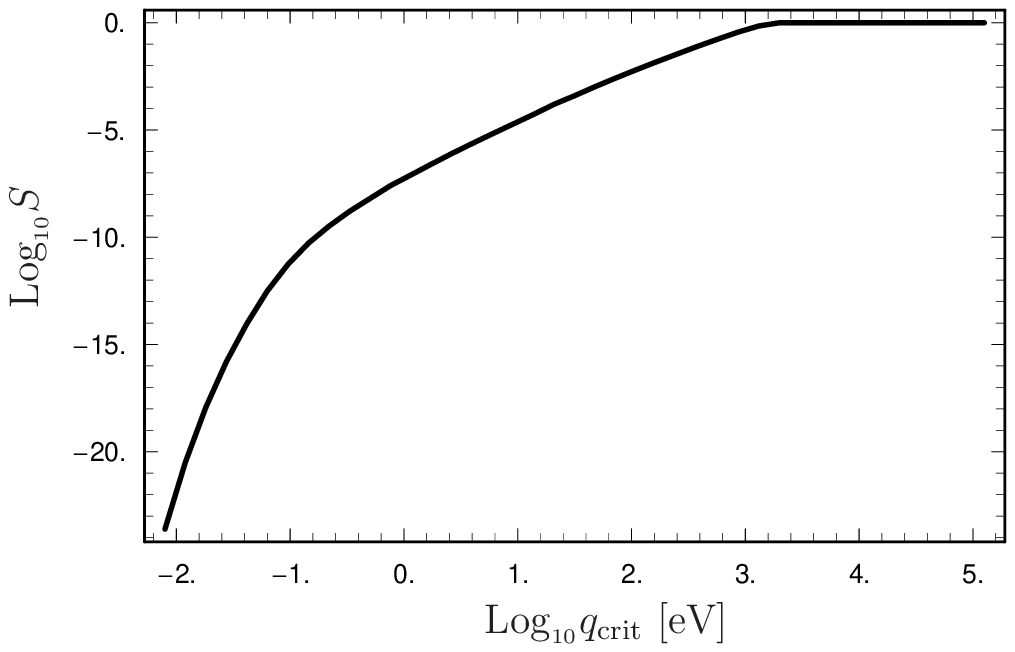}   
    \includegraphics[width=7.5cm]{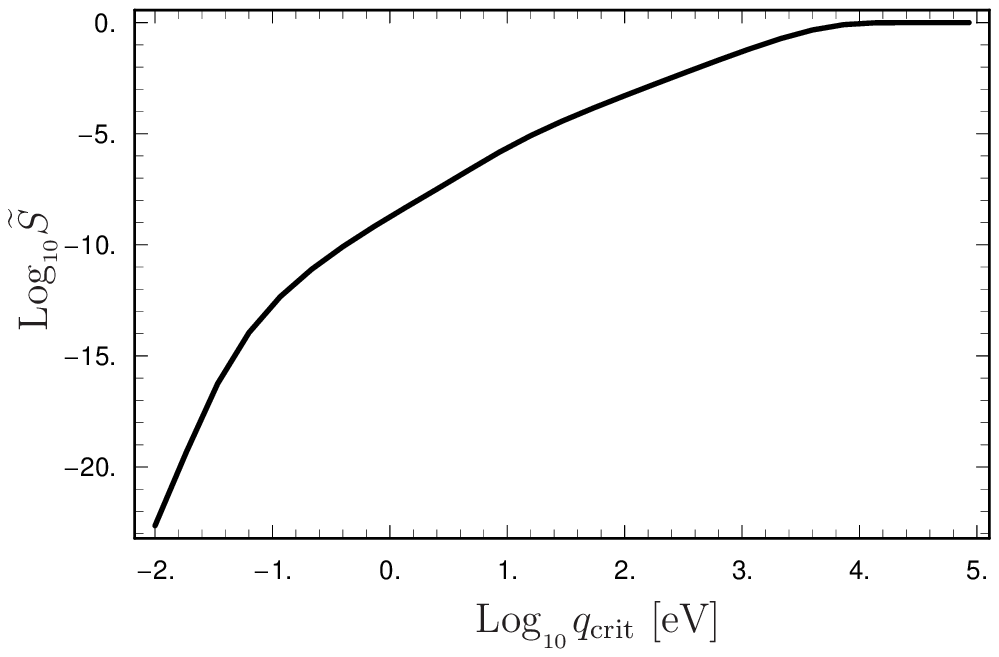}   
    \caption{\label{SUPGAINq2}Suppression factor $S$ for CAST, and $\tilde{S}$ for the energy loss
arguments as a function of
$q_{\rm{crit}}$.}
\end{figure}

Proceeding along the lines of the previous section we can calculate the suppression factors for the
CAST experiment $S$ and the corresponding $\tilde{S}$ that appears in the energy loss considerations.
The results are plotted in Fig.~\ref{SUPGAINq2}.

Using the required suppression \eqref{castfactor}, $S\sim 10^{-20}$, for CAST and
\eqref{lossfactor}, $\tilde{S}\sim 10^{-10}$, for the energy loss arguments, we infer
that sufficient suppression requires
\begin{equation}
q_{\rm{crit}}\lesssim 10^{-2}\, \rm{eV}.
\end{equation}
Although this seems rather small it is nevertheless quite big compared to the typical momentum
transfer in the PVLAS experiment,
\begin{equation}
q_{\rm{PVLAS}}\approx \frac{m^{2}_{\phi}}{2\omega}\sim 6\times10^{-7}\,\rm{eV}.
\end{equation}

\subsection{Kinematical Suppression}

So far, we have suppressed the production of ALPs by reducing their coupling to
photons. Now, we consider the possibility that the suppression originates from
an increase of the ALP's effective mass. Clearly, if the latter is larger than
the temperature, only the Boltzmann tail of photons with
energies higher than the mass can contribute to ALP production.

If we consider macroscopic environmental parameters $\eta(R)$ and, again, assume the simplest dependence on
these parameters,
\be
\label{massdependence}
m(\eta < \eta_{\rm{crit}})= m\ \ (\sim \textrm{meV}), \hspace{1cm} m(\eta>\eta_{\rm{crit}})=\infty, \ee
the suppression is identical to the one computed in Sect. \ref{macroscopic}, since the Boltzmann tail vanishes
for infinite mass.
Accordingly,
Figs.~\ref{fig4} and  \ref{totalflux1}
give the correct suppression also for the case of an environment dependent mass.

Before we continue let us point out that a strong dependence of the mass on environmental
parameters such as in Eq. \eqref{massdependence} is problematic because it requires a strong
coupling between the ALP and its environment. This still holds even if we require only
$m(\eta>\eta_{crit})\gtrsim 10\,\rm{keV}$. The strong coupling is likely to lead to
unwanted side effects
, as we commented in Sec. II, but let us however discuss some phenomenological aspects which could
distinguish kinematical suppression from a dynamical suppression via the coupling.
As an explicit example, we discuss a dependence on the density $\rho$. The wave equation for the
ALP will be
\be \label{waveequation} \Box\phi + m^{2}(\rho(x))\phi=0 . \ee
The effective mass,  $m(\rho(x))$, acts as a potential for $\phi$. This can actually lead to a new
way to avoid the CAST bound. For example consider a situation where ALPs are emitted with energy
$\omega$. When they encounter a macroscopic ``wall'' with $m(\rho_{\rm{wall}})>\omega$ on their way
to the CAST detector, they will be reflected due to energy conservation (tunneling through a
macroscopic barrier is negligible). In other words, they will not be able to reach the CAST
detector and can not be observed. In this case only the energy loss arguments require a suppression
of the production \eqref{lossfactor} whereas the stronger constraint \eqref{castfactor} from CAST
is circumvented by the reflection.

\begin{figure}
\begin{center}
\includegraphics[width=12cm]{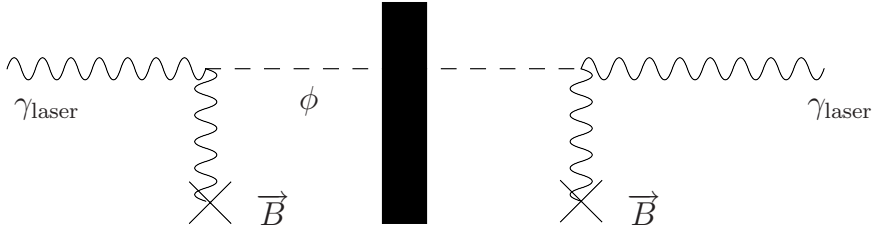}
\end{center}
\vspace{-1cm} \caption{Schematic view of a ``light shining through a wall'' experiment.
(Pseudo-)scalar production through photon conversion in a magnetic field (left), subsequent travel
through an (opaque)  wall, and final detection through photon regeneration
(right). \hfill \label{fig:ph_reg}}
\end{figure}

This effect will also play a central role in the interpretation of the PVLAS result
in terms of an ALP. Note that the interaction region (length $L$) of the PVLAS set up is located inside a
Fabri-Perot cavity which enlarges the optical path of the light inside the magnetic field by a
factor $N_r\sim 10^5$ accounting for the number of reflections inside the cavity..
In the standard ALP scenario, the ALPs created along one path cross the
mirror and escape from the cavity. Coherent production takes place only over the length $L$. The net result produces a rotation non-linear in $L$ but only linear in $N_r$
\cite{Raffelt:1987im},
\begin{equation}
|\Delta\theta|=N_{r}\left(\frac{B\omega}{M m^2}\right)^{2}\sin^{2}\left(\frac{L m^2}{4\omega}\right).
\end{equation}
However, if $m=m(\rho)$ the ALPs have a potential barrier in this mirror and
they will be reflected in the same way as the photons.
In fact, the whole setup now acts like one pass through an interaction region of length $N_{r}L$.
The ALP field in the cavity will increase now non-linearly in $N_{r}L$  modifying the
predicted rotation in the following way
\begin{equation}
\label{modified}
|\Delta\theta_{\rm modified}|=\left(\frac{B\omega}{M m^2}\right)^{2}
\sin^{2}\left(\frac{N_{r} L m^2}{4\omega}\right),
\end{equation}
where $\omega$ is the frequency of the laser. For small enough $m\lesssim {\rm
few}\times10^{-6}\,\rm{eV}$ this grows as
\begin{equation}
|\Delta\theta_{\rm modified}|\approx \frac{N^{2}_{r}L^{2}B^{2}}{16 M^{2}}.
\end{equation}
Under these conditions the PVLAS experiment cannot fix $m$ using the exclusion bounds from BFRT.
Using Eq. \eqref{modified} the rotation measurement \textit{suggests}, however, a much more
interesting value
\begin{equation}
M_{\rm modified} \sim 10^8\ {\rm GeV},\quad{ \rm{for} }\quad m\lesssim {\rm few}\times 10^{-6} {\rm{eV}},
\end{equation}
where we have used $l\sim 1$ m, $N_r\sim 10^5$ and $\omega\sim 1$ eV for the PVLAS setup.
That could be reconciled more easily with astrophysical bounds within our framework.

Such an effective mass will also play a role in  ``light shining through a wall'' experiments (cf.
Fig.~\ref{fig:ph_reg}). Typically, the wall in such an experiment will be denser than the critical
density $\rho_{\rm{crit}}$ required from the energy loss argument. Consequently, an ALP produced on
the production side of such an experiment will be reflected on the wall and cannot be reconverted
in the detection region. Hence, such an experiment would observe nothing if a density dependent
kinematical suppression is realized in nature.

\section{\label{summary}Summary and conclusions}

The PVLAS collaboration has reported a non-vanishing rotation of the polarization of a laser beam
propagating through a magnetic field. The most common explanation for such a signal would be the
existence of a light (pseudo-)scalar axion-like particle (ALP) coupled to two photons. However, the
coupling strength required by PVLAS exceeds astrophysical constraints by many orders of magnitude.
In this paper, we have quantitatively discussed ways to evade the astrophysical bounds by
suppressing the production of ALPs in astrophysical environments, in particular in the Sun.

The simplest way to suppress ALP production is to make the coupling $1/M$ of ALPs to photons small
in the stellar environment. Motivated by microphysical
models~\cite{Masso:2005ym,Masso:2006gc,Abel:2006qt,Mohapatra:2006pv}, we considered a dependence of
$M$ on environmental parameters, such as temperature, plasma mass $\omega_P$, or density $\rho$.
One of our main results is that it is not sufficient to suppress production in the center of the
Sun only. One has to achieve efficient suppression also over a significant part of the more outer
layers of the Sun. As apparent from Tables~\ref{PVLAS_SUN_HB}, \ref{tab2} and
Eq.~\eqref{castfactor}, it is possible to reconcile the PVLAS result with the bound from the CERN
Axion Solar Telescope (CAST) if strong suppression sets in at sufficiently low critical values of
the environmental parameters, e.g. $\rho\sim 10^{-3}$~g/cm$^3$, or $\omega_P\sim $~eV. The bounds
arising from solar energy loss considerations are less restrictive (cf. Eq.~\eqref{lossfactor} and
Figs.~\ref{fig2}, \ref{totalflux1}). As an alternative suppression mechanism, we have also
exploited an effective mass that grows large in the solar environment. This case, too, requires
that the effect sets in already for low critical values of the environmental parameters (cf.
Figs.~\ref{fig4}, \ref{totalflux1}).

Most proposed near-future experiments to test the PVLAS ALP interpretation are of the ``light
shining through a wall'' type (cf. Fig.~\ref{fig:ph_reg}). In these experiments, the environment,
i.e. the conditions in the production and regeneration regions, may be modified. The above
mentioned critical values are small enough that they may be probed in such modifications. For
example, a density dependence may be tested by filling in buffer gas.

In conclusion, the PVLAS signal has renewed the interest in light bosons coupled to photons. The
astrophysical bounds, although robust, are model-dependent and may be
relaxed by many orders of magnitude. Therefore, the upcoming laboratory experiments are very
welcome and may well lead to exciting discoveries in a range which was thought to be excluded.

\section{acknowledgments}
Two of us (EM and JR) acknowledge support by the projects FPA2005-05904 (CICYT) and 2005SGR00916
(DURSI).

\end{document}